\documentclass[sigconf]{acmart}
\AtBeginDocument{%
  }

\setcopyright{acmlicensed}
\copyrightyear{2026}
\acmYear{2026}
\acmDOI{XXXXXXX.XXXXXXX}
\acmConference[SURE '26]{Software Understanding and Reverse Engineering}{November 15--19, 2026}{The Hague, The Netherlands}
\acmISBN{}

\usepackage{multirow}
\usepackage{todonotes}
\usepackage[linesnumbered,ruled,vlined]{algorithm2e}
\newcommand{\cmark}{\ding{51}} 
\newcommand{\xmark}{\ding{55}} 

\newcommand{\name}{{\textsc{Chisel}}\xspace}

\begin{document}



\title{\name-ing Back Source Code with AI-enabled Iterative Recovery}


\author{Varun Kohli}
\correspondingauthor
\affiliation{%
  \institution{A*STAR Institute of Advanced Intelligence and Computing}
  \country{Singapore}
  }
\email{Kohli\_Varun@a-star.edu.sg}

\author{N Raghava}
\affiliation{%
  \institution{Coimbatore Institution of Technology}
  \city{Coimbatore}
  \country{India}
  }
\email{raghava.research.cs@gmail.com}

\author{Biplab Sikdar}
\affiliation{%
  \institution{National University of Singapore}
  \country{Singapore}
  }
\email{bsikdar@nus.edu.sg}

\author{Dinil Mon Divakaran}
\affiliation{%
  \institution{A*STAR Institute of Advanced Intelligence and Computing}
  \country{Singapore}
  }
\email{Dinil\_Divakaran@a-star.edu.sg}


\begin{abstract}
Decompilation aims to recover high-level, compilable, and semantically equivalent code from binaries. Traditional decompilers produce pseudo-C that is difficult to read and does not compile, while the recent LLM-assisted approaches generate readable, but semantically incorrect code. LLM-aided iterative recovery is an emerging branch of research, but prior works rely on supplied test suites for semantic recovery. In this work, we present \name, a test suite-free framework to iteratively recover source code from Ghidra-derived pseudo-C. \name uses simple yet effective feedback from a compiler (static analysis) and a coverage-guided fuzzer (differential analysis), augmented by rich observables for grounded divergence detection and feedback, cross-iteration divergence memory, and best candidate retention. We systematically evaluate \name for compilation and semantic recovery, feedback oracle soundness, and iteration overhead on 120 ExeBench functions compiled for the x86-64 architecture, across four optimizations (\texttt{O0-O3}), in both stripped and unstripped variants, using the open-weight Gemma4:31b LLM. \name, with all recommended features, achieves an average of 96.1\% re-compilability and 79.8\% re-executability rates at an average of 2.1 iterations. Significantly, \name recovers 26\% of first-generation execution errors. At the same time, \name's feedback oracle falsely accepts only 9.4\% candidates. Lastly, \name performs significantly better than two recent prior work on LLM-assisted decompilation. 
\end{abstract}

\ccsdesc[500]{Security and privacy~Software reverse engineering}


\received{30 June 2026}
\received[revised]{28 August 2026}
\received[accepted]{28 August 2026}

\maketitle

\section{Introduction}
Decompilation, i.e., the recovery of high-level language from compiled code, is valuable for reverse engineering, malware analysis, vulnerability detection and mitigation, analyzing patches and auditing legacy and firmware binaries for which no source is available~\cite{SOK-offensive-binary-2016, ahoy2024}. The recovery of readable, compilable, and semantically faithful source from machine code is a challenging task, and there has been significant effort in developing faithful decompilers that let an analyst read, recompile, patch, and re-deploy a binary for which no source is available. Traditional decompilers such as Ghidra~\cite{ghidra} and IDA Pro~\cite{idapro} reconstruct control flow and data types using hand-built heuristics, which is difficult to read, rarely compilable, and does not guarantee semantic fidelity. The resultant pseudo C is typically obscured by synthetic types, machine-level artifacts and Application Binary Interface (ABI)-level scaffolding. This has led to the development of AI and LLM-based approaches to translate assembly or decompiler output into readable and compilable source code~\cite{10646727,tan2024llm4decompile,hu2024degpt, gao2025decompilebench, basque2026decompiling}, which can be broadly classified into two categories: \textit{Non-iterative decompilers} use finetuned LLMs to recover source code, relying solely on the LLM's intrinsic capability to generate top-1 or top-k candidates without providing feedback for improvement after failed generations~\cite{tan2024llm4decompile,jiang2025nova,armengol2024slade,dramko2026idioms}. \textit{Iterative decompilers} address these limitations` by using feedback derived from in-loop oracles to guide the LLM towards better generation~\cite{hu2024degpt,wong2025decllm,cui2026pcodetrans,zhang2026constraint}. 

However, several methods rely on native test suites to identify semantic divergence that is then passed to the LLM as feedback \cite{wong2025decllm, cui2026pcodetrans}. This cannot be assumed when reversing binaries derived from unknown source code, and is the biggest limitation of prior work. These works also use a sanitizer to fix memory corruption errors which is not faithful when done asymmetrically only to the candidate, since the original source may also showcase such properties. Some works use symbolic execution as part of a finetuning pipeline or for multi-role generation, but consequently suffer from path explosion, making it challenging to generalize across control-flow types~\cite{state-aware-SE-USENIX-2022, zou2025d, hu2024degpt}. Further, in-loop feedback coverage is typically limited to compiler errors and return value divergence~\cite{wong2025decllm, zhang2026constraint}.

\begin{table*}[t]
\centering
\caption{A comparison of LLM-assisted decompilation methods related to \name.}
\label{tab:related}
\setlength{\tabcolsep}{3pt}
\renewcommand{\arraystretch}{1.1}
\resizebox{\linewidth}{!}{%
\begin{tabular}{l|c|c|ccc|c|c|c|c|c}
\hline
\multirow{2}{*}{\textbf{Method}} & \multirow{2}{*}{\textbf{Iterative}} & \multirow{2}{*}{\textbf{test suite-free}} & \multicolumn{3}{c|}{\textbf{In-loop Oracle}}  & \multirow{2}{*}{\textbf{\begin{tabular}[c]{@{}c@{}}Soundness\\ Evaluation\end{tabular}}} & \multirow{2}{*}{\textbf{\begin{tabular}[c]{@{}c@{}}Signature Type\\ Extraction\end{tabular}}} & \multirow{2}{*}{\textbf{\begin{tabular}[c]{@{}c@{}}Rich\\ Observables\end{tabular}}} & \multirow{2}{*}{\textbf{\begin{tabular}[c]{@{}c@{}}Divergence\\ Memory\end{tabular}}} & \multirow{2}{*}{\textbf{\begin{tabular}[c]{@{}c@{}}Best Candidate\\ Retention\end{tabular}}} \\ \cline{4-6}
 &   &         & \multicolumn{1}{c}{Compiler}              & \multicolumn{1}{c}{Sanitizer}             & Fuzzer &         & &        &      &              \\ \hline
LLM4Decompile \cite{tan2024llm4decompile}  & \xmark & \cmark    & \multicolumn{1}{c}{\xmark} & \multicolumn{1}{c}{\xmark} & \xmark & \xmark  & \xmark         & \xmark & \xmark              & \xmark        \\
Nova \cite{jiang2025nova}           & \xmark & \cmark    & \multicolumn{1}{c}{\xmark} & \multicolumn{1}{c}{\xmark} & \xmark & \xmark  & \xmark         & \xmark & \xmark              & \xmark        \\
Idioms \cite{dramko2026idioms}         & \xmark & \cmark    & \multicolumn{1}{c}{\xmark} & \multicolumn{1}{c}{\xmark} & \xmark & \xmark  & \xmark         & \xmark & \xmark              & \xmark        \\
DeGPT \cite{hu2024degpt}          & \xmark & \cmark    & \multicolumn{1}{c}{\xmark} & \multicolumn{1}{c}{\xmark} & \xmark & \xmark  & \xmark         & \xmark & \xmark              & \xmark        \\
DecLLM \cite{wong2025decllm}         & \cmark & \xmark    & \multicolumn{1}{c}{\cmark} & \multicolumn{1}{c}{\cmark} & \cmark & \xmark  & \xmark         & \xmark & \xmark              & \xmark        \\
PCodeTrans \cite{cui2026pcodetrans}     & \cmark & \xmark    & \multicolumn{1}{c}{\cmark} & \multicolumn{1}{c}{\cmark} & \xmark & \xmark  & \xmark         & \cmark & \xmark              & \xmark        \\
Agent4Decompile \cite{zhang2026constraint}& \cmark & \xmark    & \multicolumn{1}{c}{\cmark} & \multicolumn{1}{c}{\xmark} & \xmark & \xmark  & \xmark         & \xmark & \xmark              & \xmark        \\ \hline
\textbf{\name} (ours)   & \cmark & \cmark    & \multicolumn{1}{c}{\cmark} & \multicolumn{1}{c}{\xmark} & \cmark & \cmark  & \cmark         & \cmark & \cmark              & \cmark        \\ \hline
\end{tabular}
}
\end{table*}

To address the above limitations, we develop and present \name\footnote{\underline{C}ontrolled \underline{H}euristics for \underline{I}terative \underline{S}emantic \underline{E}xtraction and \underline{L}ifting}, a test suite-free iterative enhancement decompiler. We make the following contributions: 
\begin{enumerate}
  \item A lightweight and test-suite-free iterative decompilation framework whose in-loop signals are derived only from the original binary using a compiler and coverage-guided fuzzer, aided by rich observables for grounded differential analysis, cross-iteration divergence memory, and best-candidate retention. Our code is available on GitHub\footnote{https://github.com/VarunKohli18/chisel}.
  \item A systematic six-arm ablation of \name's iteration harness, building up from single-shot LLM generation to our end-to-end design, highlighting the incremental impact of each components across three axes of evaluation, namely \textit{recovery} of source, which covers Re-compilability (RC), Re-executability (RE), regression, and three new recovery metrics that we define in Section \ref{sec:experiments} to evaluate the effectiveness of an iterative decompiler; \textit{soundness} of the feedback oracle, which covers the oracle's False-acceptance (FA) and False-rejection (FR) rates judged against the ground-truth test suite; and \textit{overhead}, which includes the average number of iterations.
  \item A systematic evaluation on 120 ExeBench functions \cite{armengol2022exebench} across four optimization levels (\texttt{O0-O3}), with both stripped and unstripped x86-64 binaries, using an off-the-shelf, open-weight Gemma4:31b model in all experiments, that reports 96.1\% RC, 79.8\% RE, 9.4\% FA, and 26.2\% recovery after first-generation failure. We also report better performance than LLM4Decompile \cite{tan2024llm4decompile} and Agent4Decompile \cite{zhang2026constraint}. 
\end{enumerate}

\section{Related Work}
\label{sec:related}

Table~\ref{tab:related} compares related works over methodologies and design. The first family of LLM-assisted decompilers produces candidate source code without behavioral feedback to improve generation. LLM4Decompile~\cite{tan2024llm4decompile} and Nova~\cite{jiang2025nova} fine-tune LLMs to translate assembly or Ghidra pseudo-C into compilable and re-executable C, and Idioms~\cite{dramko2026idioms} jointly recovers code and its user-defined types from decompiler output. These methods do not utilize feedback derived from the generated code or differential analysis to improve the quality of generation, relying only on the LLMs inherent capability to recover good-quality source code. DeGPT~\cite{hu2024degpt} improves readability with a multi-role pipeline and a static symbolic check that refuses edits that change the function.

Iterative decompilation methods overcome the limitations of non-iterative approaches through feedback loops. DecLLM~\cite{wong2025decllm} repairs compiler errors, identifies semantic divergence using coverage-guided AFL++~\cite{fioraldi2020afl} seeded by a supplied ground-truth test suite, and employs a sanitizer on the generated candidate to fix memory misuse. PCodeTrans~\cite{cui2026pcodetrans} validates hot-swapped function using the program's official test suite, then localizes failures with breakpoint-matched differential tracing that compares control flow and variable values against the original. Both works use compilers and sanitizers, while DecLLM also uses a fuzzer for divergence feedback. However, the use of a sanitizer to asymetrically fix the generated code (while not the origin) is unfaithful decompilation. Agent4Decompile~\cite{zhang2026constraint} proposes a three-tier hierarchical agentic approach covering syntax, compilation, and execution checks that assume the availability of supplied test cases, showcasing the first agentic approach to pseudocode enhancement. 

The main limitation of the above methods is the reliance on a test suite for in-loop feedback, which may not be possible to access when reversing binaries whose source is unknown. Additionally, with the exception of PCodeTrans \cite{cui2026pcodetrans}, prior works typically have limited coverage of observable behaviors for differential analysis, and show limited capability. In contrast, \name is an iterative enhancement framework that improves the output of traditional decompilers without supplied test suites. It uses in-loop signals derived solely from the original binary, performing static analysis using a compiler for compilation failures and differential analysis through type-aware coverage-guided fuzzing over a rich observable space that includes return values, output buffer contents, stdout, exit codes, timeouts, and crash signals. \name also integrates cross-iteration divergence memory and best candidate retention if the feedback oracle doesn't accept any candidate after the last iteration.

\section{Proposed Framework: \name}
\label{sec:proposed}
This section presents the design components of \name, including the iterative enhancement algorithm, constituent oracles, features, prompts and parameter selection.

\subsection{Iterative Enhancement Algorithm}
\begin{figure}[t]
    \centering
    \includegraphics[width=0.8\linewidth]{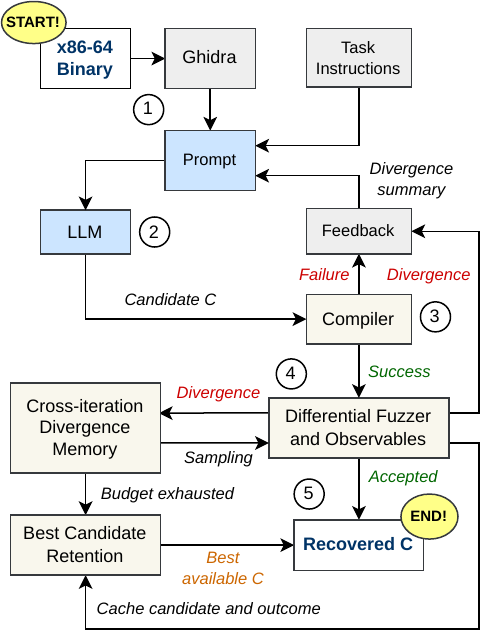}
    \caption{Overview of \name. The overall sequence of steps is marked with numbers.}
    \label{fig:chisel}
\end{figure}

\name's novelty lies in its test-suite-free iterative refinement through compiler and type-aware differential fuzzer feedback, augmented by features for broader observability, divergence memory, and best candidate retention. Fig.~\ref{fig:chisel} provides an overview of \name and its iterative procedure is depicted by Algorithm~\ref{alg:loop}. The algorithm takes Ghidra pseudo-C ($P$) of a single function, the original reference object ($o$) that needs to be decompiled, iteration budget ($K$), feedback oracles ($\mathcal{O}$), feedback budgets ($d_c,d_f$), and features ($\mathcal{F}$). The algorithm returns one compilable C function \texttt{func0}. The subsequent text explains the algorithm assuming all components are active.

Each round samples a single candidate $c$ from the LLM at zero temperature, using a self-contained prompt assembled from $P$, the previous candidate ($\textit{prev}$), and the current feedback ($f_b$) derived from oracles (line 2-22): $\mathcal{O}\cdot$\textsc{compiler} is a static feedback oracle that compiles the generated C, parses observed failures, and shares them as $f_b\leftarrow\textsc{Diagnostics(c)}$ (lines 8-9), nudging the LLM towards generating compilable C in the next iteration. If $c$ compiles, $\mathcal{O}.$\textsc{differential} searches for inputs that expose divergence of $c$ relative to $o$ through coverage-guided entropic libFuzzer mining of $B$ samples from a type-aware seed corpus (\textsc{TypeSeeds}) derived from the parsed signature ($\sigma$) for a better starting point than random (lines 11-16). Discriminating inputs are sampled from the accumulated set (E) and passed as feedback $f_b\leftarrow\textsc{Sample}(E)$ (lines 17-21). Only a maximum of $d_c$ compilation errors and $d_f$ fuzzer divergences are passed as feedback and the remaining are appended as a count instead to stay within token budgets. If neither oracle flags an issue, $c$ is returned as the recovered source (lines 23-24).

\begin{algorithm}[t]
\caption{\name's iterative enhancement loop.}
\label{alg:loop}
\KwIn{pseudo-C $P$, reference object $o$, round budget $K$, feedback budgets $d_c,d_f$, oracles $\mathcal{O}$, features $\mathcal{F}$}
\KwOut{a C function \texttt{func0}}
$\textit{prev}, f_b, M, \mathcal{C} \gets \varnothing$;\

\For{$k \gets 1$ \KwTo $K$}{
  
  $c \gets \textsc{LLM}(P, \textit{prev}, f_b)$;\
  
  \If{$\mathcal{O} = \varnothing$}{
  
  \Return $c$ \tcp*{Single-shot generation}
  }
  
  $\textit{prev} \gets c$;\ $f_b \gets \varnothing$;\ 
  
  $\textit{accept} \gets \texttt{true}$;\
  
  \If{$\mathcal{O}.$\textsc{compiler} \textbf{and} $\neg\,\textsc{Compile}(c)$}{
    
    $f_b \gets \textsc{Diagnostics}(c)$;\
    
    \textbf{continue}\;
    }
    
  \If{$\mathcal{O}.$\textsc{differential}}{
  
    $\sigma \gets \textsc{ParseSignature}(c)$;\

    \eIf{$\mathcal{F}.\textsc{observe}$}{

    $X \gets \textsc{libFuzzer}(c,o,\sigma,B,\textsc{TypeSeeds}(\sigma))$;\
    }{

    $X \gets \textsc{libFuzzer}(c,o,\sigma,B,\textsc{RandomSeeds}(\sigma))$;\
    }
    
    $E \gets \textsc{Diff}(c, o, X \cup M)$;\
    
    \If{$\mathcal{F}.\textsc{memory}$}{
    
    $M \gets M \cup E$ \tcp*{Memory accumulation}
    }
    
    \lIf{$E \neq \varnothing$}{
    
    $f_b \gets \textsc{Sample}(E)$;\ 
    
    $\textit{accept} \gets \texttt{false}$
    }
  }
  \If{$\textit{accept}$}{
  
  \Return $c$ \tcp*{Accepted candidate}
  }
  
  $\mathcal{C} \gets \mathcal{C} \cup \{c\}$;\
}

\If{$\mathcal{F}.\textsc{retain\_best}$}{

\Return $\arg\min\textsc{Divergence}(\mathcal{C})$ \tcp*{Best candidate}
}{

\Return $c$
}

\end{algorithm}

Beyond the oracles, \name is equipped with three key \textit{features} to improve recovery of failed generations and oracle soundness. (1)~$\mathcal{F}\cdot$\textsc{observe} (lines 13-17) makes the differential analysis type-aware and extends observation to return values, output buffer contents, stdout, timeouts, exit codes, and crash signals, and compares pointer results by content rather than raw address, providing a richer differential feedback to the LLM for improvement. (2)~$\mathcal{F}.$\textsc{memory} (lines 18-19) grows $M$ with the deduplicated divergence-inducing inputs across iterations and replays them under the same sampling budget to ensure that the generation maintains performance on prior divergences and converges.  Finally, to handle no accepted generation or the regression of code quality through iterations, (3) $\mathcal{F}.$\textsc{retain\_best} (lines 23-24) returns the best-performing candidate from the loop as the recovered source code. Overall, \name provides a novel iterative harness that nudges generation towards compilable and functionally similar code.

\begin{table*}[t]
\centering
\caption{Prompt components and descriptive samples.}
\label{tab:prompts}
\renewcommand{\arraystretch}{1.1}
\setlength{\tabcolsep}{6pt}
\begin{tabular}{|p{0.15\linewidth}|p{0.81\linewidth}|}
\hline
\textbf{Component} & \multicolumn{1}{c|}{\textbf{Content}} \\ \hline
Task instruction &
\texttt{\scriptsize``Below is Ghidra pseudo C for one function. Rewrite it as compilable C named func0 that
reproduces the original's behavior exactly, the same return value and the same writes through its
pointer arguments. Keep the computation and control flow as written: preserve the order of
operations and do not restructure or simplify loops, reorder side effects, or add or remove
returns. But omit the decompiler scaffolding that reflects the machine and the ABI rather than the
program, such as stack-protector canary checks, the placeholders the decompiler invents for values
it could not recover, and register or stack spill temporaries, and rewrite synthetic types and raw
memory addressing as ordinary C. The guiding rule: keep every statement that affects the return
value or bytes written through the pointer arguments, and drop every statement that does not.
Output only the C source of func0, no prose and no comments, and include every header and type it
needs.''} \\ \hline
Ghidra pseudo-C & \texttt{\scriptsize``Ghidra pseudo C: \{PSEUDO\_C\}''} \\ \hline
Previous generation & \texttt{\scriptsize E.g.``Previous attempt: \{PREVIOUS\}''} \\ \hline
Compiler feedback &
\texttt{\scriptsize E.g.\ 1.``Compilation failed: \{ERRORS\}. Fix it.''} \newline
\texttt{\scriptsize E.g.\ 2.``Compilation timed out. Simplify the control flow and try again.''} \\ \hline
Fuzzer feedback &
\texttt{\scriptsize E.g.\ 1.``On func0(...): outputs agree on the first \{COUNT\} value(s), then differ at index
\{INDEX\}: original \{ORIGINAL\_VALUES\}.''} \newline
\texttt{\scriptsize E.g.\ 2.``On func0(...): original printed \{OUTPUT\}, candidate crashed with signal \{S\}.''} \newline
\texttt{\scriptsize E.g.\ 3.``On func0(...): original printed \{OUTPUT\}, candidate timed out.''} \newline
\texttt{\scriptsize E.g.\ 4.``On func0(...): original printed \{OUTPUT\}, candidate printed \{OUTPUT'\}.''} \\ \hline
\end{tabular}
\end{table*}

\subsection{Prompts and Parameters}
Every round is driven by one self-contained prompt which is depicted in Table \ref{tab:prompts}. It consists of a fixed task instruction that guides the LLM to rewrite the pseudo C as compilable C named \texttt{func0} that reproduces the original's behavior exactly, preserving the control flow and the order of operations while dropping decompiler scaffolding that reflects the machine and the ABI rather than the program. The first round shows only this instruction and the pseudo C. Later rounds append $prev$ and $f_b$ as per availability.

We set $K=5$, the LLM is set at temperature $0$ for reproducibility, and reasoning is turned off. Candidates are compiled at \texttt{O0} optimization by default for static and differential analysis. Each round tests $c$ against $o$ on a 200-sample corpus of type-aware seeds and $B=2000$ libFuzzer-mined inputs, with a per-input timeout of $2$ CPU seconds to catch runaway loops. We set $d_c=5$ and $d_f=10$.

\section{Experimental Setup}
\label{sec:experiments}
This sections provides details of the experimental setup including hardware, software, dataset, experiments, evaluation metrics, and baselines compared. 

\subsection{Hardware, Software, and Dataset}
We run all experiments on a x86-64 Linux host with 4 NVIDIA H200 GPUs. We use the Ollama-provisioned \texttt{Gemma4:31b}\footnote{https://ollama.com/library/gemma4:31b} at int-4 quantization for our experiments. The input input and output context limits set to 24,576 and 8,192 tokens respectively. The harness uses Ghidra for lifting assembly to pseudo-C, \texttt{gcc -c} for compiler feedback, and libFuzzer for divergence feedback. We draw the 150 longest real-world functions from ExeBench's real and val-real splits \cite{armengol2022exebench}, covering applications including and not limited to, digital signal processing, cryptography, geometry, linear algebra, and data structure. The functions are given a ground truth test suite of 1000 input-output pairs using a generator and further filtered to 120 functions based on the availability of at least 200 functional test cases. The selected functions have 46-265 lines-of-code, and are compiled on x86-64 at \texttt{O0-O3} optimizations in stripped and unstripped variants. The cached dataset is available on our GitHub.

\begin{table*}[t]
\centering
\renewcommand{\arraystretch}{1.15}
\caption{Metrics derived from one run of a six-arm ablation, averaged over all variants.}
\label{tab:ablation}
\resizebox{0.9\linewidth}{!}{%
\begin{tabular}{lccccccccccl}
\toprule
\textbf{Harness Arm} & $RC$ & $RE$ & $Pass$& \textbf{$R_{comp}$} & \textbf{$R_{exec}$} & \textbf{$R_{tot}$} & $FA$ & $FR$ & \textbf{$Reg$} & $\# \bar{k}$ & \textbf{Remarks} \\
\midrule
LLM (one-shot) & 90.0 & 73.0 & 75.1 & -- & -- & -- & -- & -- & -- & 1.00 & LLM-only, no feedback \\
\texttt{+ compiler} & 96.6 & 77.1 & 79.2 & 65.6 & 0.0 & 15.1 & 20.2 & 0.0 & 0.0 & 1.18 & gcc oracle \\
\texttt{+ fuzzer} & 96.6 & 79.3 & 81.7 & 67.3 & 7.5 & 23.5 & 17.6 & 45.0 & 0.0 & 1.43 & Differential libFuzzer oracle \\
\texttt{+ observe} & 96.7 & 79.9 & 82.0 & 66.7 & 13.0 & 26.0 & 9.4 & 45.4 & 0.3 & 2.05 & Type aware and rich observables \\
\texttt{+ memory} & 96.1 & 79.6 & 81.8 & 64.1 & 14.4 & 25.9 & 9.4 & 45.1 & 0.1 & 2.06 & Divergence memory and replay \\
\texttt{+ best} (full \name) & 96.1 & 79.8 & 82.1 & 64.1 & 15.0 & 26.2 & 9.4 & 45.9 & 0.0 & 2.06 & Best-candidate retention \\
\bottomrule
\end{tabular}}
\end{table*}

\subsection{Experiments and Evaluation Metrics} 
We evaluate \name through a systematic, six-arm incremental ablation of its harness starting from a one-shot LLM, and adding oracles (compiler, fuzzer) and features (observables, cross-iteration memory, best candidate retention), where each arm retains the prior features and the final arm corresponds to the complete \name framework. The ablation enables us to evaluate the value of each component along three key axes, i.e., \emph{recovery} which measures how much correct source C the loop produces, \emph{soundness} measures whether the feedback oracle's accept and reject decisions agree with the test suite, and \emph{overhead} which measures the cost of the added complexity. All percentage metrics are computed on the single candidate the loop finally returns, scored against the held-out suite, while overhead is measured in average number of iterations ($\bar{k}$). A candidate \emph{compiles} if it builds, and \emph{re-executes} if it matches the original on every suite input. The metrics, including our newly defined $R_{comp},R_{exec},R_{tot}$, are as follows:
\begin{itemize}
\item $RC$ and $RE$: the fractions of returned candidates that compile and that re-execute (pass \underline{all} test cases), respectively.
\item $Pass$: the average fraction of test cases passed across samples. 
\item $R_{comp}$: of the samples whose first generation did not compile, the fraction whose returned candidate compiles.
\item $R_{exec}$: of the samples whose first generation compiled but did not re-execute, the fraction whose returned candidate re-executes.
\item $R_{tot}$: of all samples whose first generation did not re-execute, the fraction whose returned candidate re-executes, i.e. the loop's overall recovery from first-generation failure.
\item $FA$: of the candidates the oracle accepted, the fraction that stopped on an incorrect candidate.
\item $FR$: of the candidates the oracle did not accept, the fraction that were labeled as correct by the evaluation test suite.
\item $Reg$: of all samples whose first generation already re-executed, the fraction whose returned candidates did not re-execute (regressed).
\item $\bar{k}$: the mean number of iterations the decompiler runs across all samples in the dataset.
\end{itemize}





Beyond $RC$, $RE$, and $Pass$, a good iterative decompiler is also defined by its $R_{comp},R_{exec}, R_{tot}$, and $Reg$. The feedback oracle should (ideally) only accept good-quality candidates and reject incorrect ones. Thus, the design of an oracle should report low $FA$ and $FR$. Over-rejection (high $FR$) leads to longer iterative loops and higher $\bar{k}$. However, unless the iteration budget and overhead constraints are tight, this means the oracle is strict, making it a less serious concern since the LLM is then triggered to generate ``better'' code. However, high $FA$ implies the system produces low-quality candidates, which is undesirable for any decompiler. 

\subsection{Baselines}
We evaluate two prior works on our dataset and evaluation harness: i)~LLM4Decompile-9B-v2 from~\cite{tan2024llm4decompile} and available on HuggingFace \cite{llm4decompile-9b-v2}, and ii)~Agent4Decompile \cite{zhang2026constraint} (code available at \cite{agent4decompile}). We replace the LLM in Agent4Decompile with Gemma4:31 for a fair comparison of the harness. Further, to align with our test-suite-free setting , no method has access to the ground truth test suite during iterations. Both methods use Ghidra for binary to pseudo-C lifting, and are run using their specified prompts and input formats. We do not compare with other iterative methods such as DecLLM \cite{wong2025decllm} and PCodeTrans \cite{cui2026pcodetrans} since their code is not public.

\section{Results}
\label{sec:results}

We evaluate \name's performance using metrics that capture recovery, soundness, and overhead, shown first as a six-arm ablation of its harness, then performance across optimizations for unstripped and stripped cases, and finally compare against recent work.

\begin{figure}[t]
    \centering
    \includegraphics[width=0.9\linewidth]{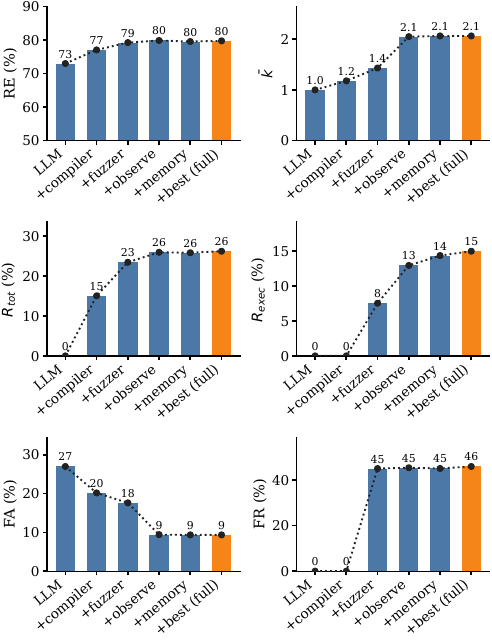}
    \caption{Key metric trends across the six-arm ablation highlighting their value in driving recovery and oracle soundness.}
    \label{fig:key-metrics}
\end{figure}

\begin{figure}[t]
    \centering
    \includegraphics[width=0.9\linewidth]{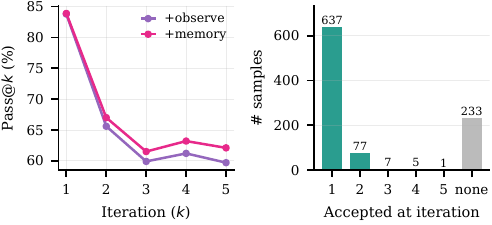}
    \caption{Per-iteration evaluation.}
    \label{fig:convergence}
\end{figure}

As Table \ref{tab:ablation} and Fig \ref{fig:key-metrics} show, the raw one-shot model compiles $90.0\%$ and re-executes $73.0\%$ of functions. Adding the compiler oracle lifts re-compilation to $96.6\%$ with $R_{comp}=65.6\%$, but is unable to repair semantic behavior after first generation ($R_{exec}=0$), since it accepts the first candidate that builds without discriminating over behavioral correctness. The differential fuzzer adds the required behavioral signal to improve $RE$ and $R_{exec}$, recovers $7.5\%$ of compiled-but-incorrect first generations, raises $RE$ to $79.3\%$, and lowers $FA$ from $27\%$ for the LLM arm and $20.2\%$ for the compiler arm, to $17.6\%$. The decisive soundness gain comes at the observe arm, which \emph{halves} $FA$ from $17.6\%$ to $9.4\%$ at equal $RE$ and $FR$, highlighting the value of rich observation that in turn leads to rich feedback and recovery. The memory and best-candidate retention arms show an improvement in $R_{exec}$ and reduction in regression. Samples derived from best-candidate retention report mean suite pass-rate increase from $50.9\%$ to $52\%$ compared to the previous arm. Fig. \ref{fig:convergence} (left) further highlights the importance of cross-iteration divergence memory. $Pass$ is higher for the \texttt{+memory} arm compared to \texttt{+observe}. Fig. \ref{fig:convergence} (right) shows the distribution of feedback oracle acceptance per iteration.

Based on our inspection of the dataset, Ghidra pseudo C, and the LLM's generation through \name, the residual $9.4\%\; FA$ is dominated by signature recovery errors predominantly in the stripped samples. The differential oracle tests each candidate against its \emph{own} recovered signature, so a candidate that mis-recovers an argument is self-consistent under the oracle yet fails the ground-truth suite that is built on the original signatures. This is visible in the strip split shown in Tables \ref{tab:chisel_unstripped} and \ref{tab:chisel_stripall}, where $FA=16.4\%$ on stripped binaries but only $1.7\%$ on unstripped, as stripping removes the type information needed by the decompiler (Ghidra) to recover the signature, which subsequently passes the error to the signature extractor used in our work (see \textsc{ParseSignature} at line 12 of Algorithm \ref{alg:loop}). $FR$ rises to nearly $45\%$ once a behavioral oracle is integrated into the system, suggesting that the oracle is stricter than the LLM-only and compiler arms. $FR$ falls monotonically with optimization regardless of stripped or unstripped symbols. We also observe that our reported $FR$ is inflated partly by the limitation of the test suite. The differential oracle outpaces the test suite due to a wider input domain (2000 samples per-iteration not considering cross-iteration memory samples). Its coverage includes divergences that were not included in the test suite.

\begin{table}[t]
\centering
\renewcommand{\arraystretch}{1.15}
\caption{\name's recovery and soundness on 480 \textbf{unstripped} x86-64 binaries (120 functions), by optimization level.}
\label{tab:chisel_unstripped}
\resizebox{\columnwidth}{!}{%
\begin{tabular}{lcccccc}
\toprule
\textbf{Optimization} & $RC$ (\%) & $RE$ (\%) & \textbf{$R_{exec}$} (\%) & \textbf{$R_{tot}$} (\%) & $FA$ (\%) & $FR$ (\%) \\
\midrule
O0 & 99.2 & 93.3 & 14.3 & 33.3 & 0.0 & 75.8 \\
O1 & 97.5 & 90.0 & 10.0 & 42.9 & 2.1 & 58.3 \\
O2 & 96.7 & 82.5 & 28.6 & 38.2 & 2.4 & 45.7 \\
O3 & 95.8 & 80.8 & 11.1 & 32.4 & 2.4 & 44.7 \\
\midrule
\textbf{Average} & \textbf{97.3} & \textbf{86.7} & \textbf{17.9} & \textbf{36.6} & \textbf{1.7} & \textbf{55.4} \\
\bottomrule
\end{tabular}}
\end{table}

\begin{table}[t]
\centering
\renewcommand{\arraystretch}{1.15}
\caption{\name's recovery and soundness on 480 \textbf{stripped} x86-64 binaries (120 functions), by optimization level.}
\label{tab:chisel_stripall}
\resizebox{\columnwidth}{!}{%
\begin{tabular}{lcccccc}
\toprule
\textbf{Optimization} & $RC$ (\%) & $RE$ (\%) & \textbf{$R_{exec}$} (\%) & \textbf{$R_{tot}$} (\%) & $FA$ (\%) & $FR$ (\%) \\
\midrule
O0 & 97.5 & 78.3 & 8.3 & 18.8 & 16.3 & 54.5 \\
O1 & 96.7 & 75.8 & 13.0 & 19.4 & 17.0 & 40.0 \\
O2 & 95.0 & 69.2 & 13.3 & 21.3 & 17.6 & 27.6 \\
O3 & 90.8 & 68.3 & 18.5 & 19.1 & 14.8 & 21.9 \\
\midrule
\textbf{Average} & \textbf{95.0} & \textbf{72.9} & \textbf{13.5} & \textbf{19.8} & \textbf{16.4} & \textbf{34.0} \\
\bottomrule
\end{tabular}}

\end{table}
\begin{table}[t]
\centering
\renewcommand{\arraystretch}{1.15}
\caption{Comparison with prior work.}
\label{tab:comparison}
\resizebox{\columnwidth}{!}{%
\begin{tabular}{lccccc}
\toprule
\textbf{Method} & $RC$ (\%) & $RE$ (\%) & \textbf{$R_{tot}$} (\%) & $FA$ (\%) & Iterations \\
\midrule
Ghidra pseudo-C & 35.0 & 28.1 & -- & -- & -- \\
LLM4Decompile-9B-v2~\cite{tan2024llm4decompile} & 63.2 & 17.0 & -- & -- & -- \\
\texttt{Gemma4:31b} & 90.0 & 73.0 & -- & -- & -- \\
+ compiler & 96.6 & 77.1 & 15.1 & 20.2 & 1.18 \\
+ fuzzer & 96.6 & 79.3 & 23.5 & 17.6 & 1.43 \\
A4D-\texttt{Gemma4:31b}~\cite{zhang2026constraint} & 87.3 & 77.7 & 15.7 & 11.0 & 2.68 \\
\hline
\textbf{\name (full)} & \textbf{96.1} & \textbf{79.8} & \textbf{26.2} & \textbf{9.4} & 2.06 \\
\bottomrule
\end{tabular}}
\end{table}

\name re-executes $86.7\%$ of unstripped and $72.9\%$ of stripped functions on average. Within each variant re-execution falls as optimization rises, as seen in both unstripped and stripped cases where $RE$ falls from $93.3\%$ at O0 to $80.8\%$ at O3, and $78.3\%$ at O0 to $68.3\%$ at O3, respectively. Meanwhile recovery rates $R_{exec}$ and $R_{tot}$ confirm that a growing share of that gap is repaired by the loop.

\paragraph*{Comparison with baselines}
Table~\ref{tab:comparison} presents the results of \name and two SOTA decompilers on the same 120 functions and test suite. The specialized one-shot decompiler, LLM4Decompile \cite{wong2025decllm} (LLM4Decompile-9B-v2), achieves only $17\%$ RE, which is below our LLM-only one-shot Gemma arm that achieves $73.0\%$ and highlights the significance of LLM capabilities for decompilation. We also evaluate Agent4Dedcompile (A4D)~\cite{zhang2026constraint} in a test-suite free setting. \name achieves performs better at a lower iteration cost. 

\section{Limitations, Risks and Ethics}
\label{sec:limitations}
Signature recovery from stripped binaries continues to be a limitation for decompilers. In addition, our experiments were conducted at 0 temperature for determinism and non-reasoning settings reduced overhead. Higher temperatures and LLM reasoning may help improve recovery through more creative and better generation, critical thinking, and response to feedback. Further, we use Ghidra pseudo-C of the original binary as the source of truth and further evaluation into decompiler resilience is required ~\cite{298204}. Additionally, I/O sampling, both as an in-loop oracle and evaluation instrument, is an approximation of functional similarity and does not indicate semantic equivalence. Lastly, while ExeBench is popularly used in decompilation research, future works may consider using newly released datasets, e.g.,~\cite{liu2026assemblagedeephistorycrossbuildbinarydataset}, to avoid the risk of testing LLMs on their training data.

On the ethical front, this research was conducted to push the frontier of reverse engineering. However, our work uses open-weight models, and our ideas could potentially be adopted by threat actors for unauthorized activities. We urge ethical use of this work. 

\section{Conclusion}
\label{sec:conclusion}
We presented \name, a test suite-free iterative decompiler that integrates static and differential analysis using compilers and coverage-guided fuzzers for compilation and execution recovery, augmented by rich observables, cross-iteration divergence memory, and best-candidate recovery. On a 120 function subset of ExeBench functions compiled to four optimizations on x86-64 with stripped and unstripped variants, \name showed $96.1\%$ RC and $79.8\%$ RE, and repairs $26.2\%$ of first-generation failures, while halving the in-loop oracle's FA from $17.6\%$ to $9.4\%$ relative to a return-only differential fuzzer. The decisive gain comes from richer observation rather than from more search. On the same subset \name far exceeds the specialized one-shot LLM4Decompile-9B-v2 model ($79.8\%$ vs $17.0\%$ RE) and outperforms Agent4Decompile in terms of RE, FA, and  overhead using the same LLM and test-suite assumptions. Our results highlight that, in addition to the LLM, the feedback harness is an important lever to improve decompilation performance.

\section*{Acknowledgement}
This research/project is supported by the National Research Foundation, Singapore, and the Cyber Security Agency of Singapore under the National Cybersecurity R\&D Programme and the CyberSG R\&D Programme Office (Award CRPO-GC2-ASTAR-001). Any opinions, findings, conclusions, or recommendations expressed in these materials are those of the author(s) and do not reflect the views of the National Research Foundation, Singapore, the Cyber Security Agency of Singapore, or the CyberSG R\&D Programme Office. 
\bibliographystyle{ACM-Reference-Format}
\bibliography{references}

@online{ghidra,
  author  = {National Security Agency},
  title   = {ghidra},
  year    = {2026},
  url     = {https://github.com/NationalSecurityAgency/ghidra},
  urldate = {2026-06-28}
}

@inproceedings{basque2026decompiling,
  title={{Decompiling the Synergy: An Empirical Study of Human-LLM Teaming in Software Reverse Engineering}},
  author={Basque, Zion Leonahenahe and Doria, Samuele and Soneji, Ananta and Gibbs, Wil and Doup{\'e}, Adam and Shoshitaishvili, Yan and Losiouk, Eleonora and Wang, Ruoyu and Aonzo, Simone and others},
  booktitle={Proc. NDSS},
  year={2026}
}

@INPROCEEDINGS{SOK-offensive-binary-2016,
  author={Shoshitaishvili, Yan and Wang, Ruoyu and Salls, Christopher and Stephens, Nick and Polino, Mario and Dutcher, Andrew and Grosen, John and Feng, Siji and Hauser, Christophe and Kruegel, Christopher and Vigna, Giovanni},
  booktitle={2016 IEEE Symposium on Security and Privacy (SP)}, 
  title={SOK: (State of) The Art of War: Offensive Techniques in Binary Analysis}, 
  year={2016},
}

@inproceedings{state-aware-SE-USENIX-2022,
author = {Shunfan Zhou and Zhemin Yang and Dan Qiao and Peng Liu and Min Yang and Zhe Wang and Chenggang Wu},
title = {Ferry: {State-Aware} Symbolic Execution for Exploring {State-Dependent} Program Paths},
booktitle = {31st USENIX Security Symposium (USENIX Security 22)},
year = {2022},
pages = {4365--4382},
month = aug
}

@online{idapro,
  author  = {Hex-Rays},
  title   = {IDA Pro: Power Disassembler, Decompiler \& Debugger},
  year    = {2026},
  url     = {https://hex-rays.com/ida-pro},
  urldate = {2026-06-28}
}

@inproceedings{gao2025decompilebench,
  title={DecompileBench: A comprehensive benchmark for evaluating decompilers in real-world scenarios},
  author={Gao, Zeyu and Cui, Yuxin and Wang, Hao and Qin, Siliang and Wang, Yuanda and Bolun, Zhang and Zhang, Chao},
  booktitle={Findings of the Association for Computational Linguistics: ACL 2025},
  pages={23250--23267},
  year={2025}
}

@inproceedings{tan2024llm4decompile,
  title={Llm4decompile: Decompiling binary code with large language models},
  author={Tan, Hanzhuo and Luo, Qi and Li, Jing and Zhang, Yuqun},
  booktitle={Proceedings of the 2024 Conference on Empirical Methods in Natural Language Processing},
  pages={3473--3487},
  year={2024}
}

@inproceedings{jiang2025nova,
  title={Nova: Generative language models for assembly code with hierarchical attention and contrastive learning},
  author={Jiang, Nan and Wang, Chengxiao and Liu, Kevin and Xu, Xiangzhe and Tan, Lin and Zhang, Xiangyu and Babkin, Petr},
  booktitle={International Conference on Learning Representations},
  volume={2025},
  pages={95905--95926},
  year={2025}
}

@inproceedings{dramko2026idioms,
  title={Idioms: A Simple and Effective Framework for Turbo-Charging Local Neural Decompilation with Well-Defined Types.},
  author={Dramko, Luke and Le Goues, Claire and Schwartz, Edward J},
  booktitle={NDSS},
  year={2026}
}

@inproceedings{armengol2024slade,
  title={Slade: A portable small language model decompiler for optimized assembly},
  author={Armengol-Estap{\'e}, Jordi and Woodruff, Jackson and Cummins, Chris and O'Boyle, Michael FP},
  booktitle={2024 IEEE/ACM International Symposium on Code Generation and Optimization (CGO)},
  pages={67--80},
  year={2024},
  organization={IEEE}
}

@article{wong2025decllm,
  title={Decllm: Llm-augmented recompilable decompilation for enabling programmatic use of decompiled code},
  author={Wong, Wai Kin and Wu, Daoyuan and Wang, Huaijin and Li, Zongjie and Liu, Zhibo and Wang, Shuai and Tang, Qiyi and Nie, Sen and Wu, Shi},
  journal={Proceedings of the ACM on Software Engineering},
  volume={2},
  number={ISSTA},
  pages={1841--1864},
  year={2025},
  publisher={ACM New York, NY, USA}
}

@inproceedings{hu2024degpt,
  title={DeGPT: Optimizing Decompiler Output with LLM.},
  author={Hu, Peiwei and Liang, Ruigang and Chen, Kai},
  booktitle={NDSS},
  year={2024}
}

@article{cui2026pcodetrans,
  title={PCodeTrans: Translate Decompiled Pseudocode to Compilable and Executable Equivalent},
  author={Cui, Yuxin and Gao, Zeyu and He, Shuxian and Qin, Siliang and Zhang, Chao},
  journal={arXiv preprint arXiv:2603.14855},
  year={2026}
}

@article{zhang2026constraint,
  title={Constraint-Guided Multi-Agent Decompilation for Executable Binary Recovery},
  author={Zhang, Yifan and Wang, Xiaohan and Zhang, Yueke and Huang, Yu and Leach, Kevin},
  journal={arXiv preprint arXiv:2604.23940},
  year={2026}
}

@article{zou2025d,
  title={D-lift: Improving llm-based decompiler backend via code quality-driven fine-tuning},
  author={Zou, Muqi and Cai, Hongyu and Wu, Hongwei and Basque, Zion Leonahenahe and Khan, Arslan and Celik, Berkay and Bianchi, Antonio and Xu, Dongyan and others},
  journal={arXiv preprint arXiv:2506.10125},
  year={2025}
}

@inproceedings{armengol2022exebench,
  title={ExeBench: an ML-scale dataset of executable C functions},
  author={Armengol-Estap{\'e}, Jordi and Woodruff, Jackson and Brauckmann, Alexander and Magalhaes, Jos{\'e} Wesley de Souza and O'Boyle, Michael FP},
  booktitle={Proceedings of the 6th ACM SIGPLAN International Symposium on Machine Programming},
  pages={50--59},
  year={2022}
}

@inproceedings{fioraldi2020afl,
author = {Andrea Fioraldi and Dominik Maier and Heiko Ei{\ss}feldt and Marc Heuse},
title = {{AFL++} : Combining Incremental Steps of Fuzzing Research},
booktitle = {14th USENIX Workshop on Offensive Technologies (WOOT 20)},
year = {2020},
month = aug
}

@inproceedings {ahoy2024,
author = {Zion Leonahenahe Basque and Ati Priya Bajaj and Wil Gibbs and Jude O{\textquoteright}Kain and Derron Miao and Tiffany Bao and Adam Doup{\'e} and Yan Shoshitaishvili and Ruoyu Wang},
title = {Ahoy {SAILR}! There is No Need to {DREAM} of C: A {Compiler-Aware} Structuring Algorithm for Binary Decompilation},
booktitle = {33rd USENIX Security Symposium (USENIX Security 24)},
year = {2024},
pages = {361--378},
month = aug
}

@INPROCEEDINGS{10646727,
  author={Pal, Kuntal Kumar and Bajaj, Ati Priya and Banerjee, Pratyay and Dutcher, Audrey and Nakamura, Mutsumi and Basque, Zion Leonahenahe and Gupta, Himanshu and Sawant, Saurabh Arjun and Anantheswaran, Ujjwala and Shoshitaishvili, Yan and Doupé, Adam and Baral, Chitta and Wang, Ruoyu},
  booktitle={2024 IEEE Symposium on Security and Privacy (SP)}, 
  title={"Len or index or count, anything but v1": Predicting Variable Names in Decompilation Output with Transfer Learning}, 
  year={2024},
  pages={4069-4087},
}

@inproceedings{298204,
author = {Muqi Zou and Arslan Khan and Ruoyu Wu and Han Gao and Antonio Bianchi and Dave (Jing) Tian},
title = {{D-Helix}: A Generic Decompiler Testing Framework Using Symbolic Differentiation},
booktitle = {33rd USENIX Security Symposium (USENIX Security 24)},
year = {2024},
pages = {397--414},
month = aug
}

@misc{liu2026assemblagedeephistorycrossbuildbinarydataset,
      title={ASSEMBLAGE-DEEPHISTORY: A Cross-Build Binary Dataset with Temporal Coverage}, 
      author={Chang Liu and Noah Fleischmann and Nicolò Altamura and Edward Raff and James Holt and Kristopher Micinski},
      year={2026},
      eprint={2605.21615},
      archivePrefix={arXiv},
      primaryClass={cs.CR},
      url={https://arxiv.org/abs/2605.21615}, 
}

@online{llm4decompile-9b-v2,
  author  = {LLM4Decompile},
  title   = {LLM4Decompile-9B-v2},
  year    = {},
  url     = {https://huggingface.co/LLM4Binary/llm4decompile-9b-v2},
  urldate = {2026-06-28}
}

@online{agent4decompile,
  author  = {Anonymous},
  title   = {Agent4Decompile Artifacts},
  year    = {2026},
  url     = {https://anonymous.4open.science/r/agent4decompile-artifacts-0F69/},
  urldate = {2026-06-15}
}

\appendix

\section{Generative AI Usage}
\label{app:genai}
We used generative AI tools, include Claude and Gemini, to (i) improve the language of the manuscript, (ii) generate the plotting code (verified) that renders figures from our experimental outputs, and (iii) writing the code for our prototype. All experiments, datasets, and numerical results were produced by our own pipeline, every figure and table was regenerated directly from the raw experimental outputs and checked against the underlying data. The authors reviewed all AI-assisted code, and take full responsibility for the accuracy, originality, and integrity of this work.

\end{document}